\documentclass[aps,prd,preprint,nofootinbib]{revtex4}
\usepackage{amsfonts,amsmath,amssymb}
\usepackage{mathrsfs}
\usepackage{graphicx}
\usepackage{multirow}
\usepackage{epsfig}
\usepackage{graphicx}
\usepackage{subfigure}
\usepackage{booktabs}
\usepackage{array}
\usepackage{tabularx}
\usepackage{slashed}
\usepackage{braket}
\usepackage{bm}
\usepackage{diagbox}

\begin{document}
\baselineskip 20pt
\title{Doubly heavy hadrons production in ultraperipheral collision}
\author{\vspace{1cm} Hao Yang$^1$\footnote[1]{yanghao2023@scu.edu.cn}, Jun Jiang$^{2}$\footnote[2]{ jiangjun87@sdu.edu.cn, corresponding author} and Bingwei Long$^{1,3}$\footnote[3]{bingwei@scu.edu.cn, corresponding author}\\}

\affiliation{
$^1$ College of Physics, Sichuan University, Chengdu, Sichuan 610065, China\\
$^2$School of Physics, Shandong University, Jinan, Shandong 250100, China\\
$^3$Southern Center for Nuclear-Science Theory (SCNT), Institute of Modern Physics, Chinese Academy of Sciences, Huizhou 516000, Guangdong, China\vspace{0.6cm}
}

\begin{abstract}
	We study the double heavy baryon $\Xi_{QQ'}$ and tetraquark $T_{QQ}$ production through photon-photon and photon-gluon fusion via ultraperipheral collisions at the LHC and FCC within the framework of nonrelativistic QCD factorization formalism. Various ion-ion collisions are taken into account, two cc(bb)-diquark configurations ($[cc(bb),{^3S_1}\mbox{-}\bar{\bm{3}}]$ and $[cc(bb),{^1S_0}\mbox{-}\bm{6}]$) and four bc-diquark configurations ($[bc,{^3S_1}\mbox{-}\bar{\bm{3}}]$, $[bc,{^3S_1}\mbox{-}\bm{6}]$, $[bc,{^1S_0}\mbox{-}\bar{\bm{3}}]$ and $[bc,{^1S_0}\mbox{-}\bm{6}]$) are considered in the calculation. Numerical results indicate that the $[cc,{^3S_1}\mbox{-}\bar{\bm{3}}]$ diquark provides dominant contribution for $\Xi_{cc}$ ($T_{cc}$) production, and a considerable number of $\Xi_{cc}$ ($T_{cc}$) can be produced. Due to the event topologies for ultraperipheral collision are very clear, the background from various QCD interactions can be suppressed, hence the experimental investigation for $\Xi_{cc}$ and $T_{cc}$ are feasible. The productions for $\Xi_{bc/bb}$ are also discussed, leaving only slightly possibility for $\Xi_{bc}$ through photon-gluon fusion with ultraperipheral collisions at the FCC.

\end{abstract}

\maketitle

\section{INTRODUCTION}
The existence of doubly heavy baryons, predicted by the quark model, has been a mystery for more than half century till the LHCb collaboration observed $\Xi_{cc}^{++}$ signal via $\Lambda_c^+K^-\pi^+\pi^+$ channel \cite{LHCb:2017iph} and further confirmed via $\Xi_{cc}^{++}\to\Xi_c^+\pi^+$ \cite{LHCb:2018pcs,LHCb:2018zpl}. And recently, doubly heavy tetraquark $T^+_{cc}(3875)$ has also been seen through $T_{cc}^+(3875)\to D^0D^0\pi^+$ at the LHCb \cite{LHCb:2021vvq}. The doubly heavy component within those hadrons indicates typically non-relativistic feature, hence they tend to stay close and form heavy-heavy diquark. In this way, the production of doubly heavy hadrons can be described by the non-relativistic QCD (NRQCD) \cite{Bodwin:1994jh}, which factorizes the production into two steps. The first step is to produce heavy-heavy diquark with given spin-color structure, e.g., $[cc,{^3S_1}\mbox{-}\bar{\bm{3}})], [cc,{^1S_0}\mbox{-}\bm{6}]$, which is perturbatively calculatable. The second step is reserved for the hadronization of the diquark though unperturbative QCD mechanism, e.g., fragmentation, which is encoded into long distance matrix elements (LDMEs) and fragmentation function.

Among the earlier theoretical investigations toward the production of doubly heavy baryons, only the color anti-triplet diquark configuration is proposed \cite{Falk:1993gb,Baranov:1995rc,Berezhnoy:1995fy,Doncheski:1995ye,Berezhnoy:1998aa}. As indicated in Ref. \cite{Ma:2003zk}, the contribution from color sextuplet holds the same level with color anti-triplet at the framework of NRQCD factorization. In this way, there are extensively studies for the production channels of $\Xi_{QQ'}$ through $e^+e^-$ \cite{Jiang:2012jt,Jiang:2013ej,Chen:2014frw}, ep \cite{Bi:2017nzv,Sun:2020mvl} and pp \cite{Chang:2006eu,Chang:2006xp} collisions, and also the indirect productions through W, Z, Higgs and top quark \cite{Zhang:2022jst,Luo:2022lcj,Niu:2019xuq,Niu:2018ycb}. Based on diquark-diquark picture, the productions for exotic tetraquark $T_{cc}$ are also investigated \cite{Chen:2011jtl,Hyodo:2012pm,Hua:2023zpa}. The above studies may provide opportunities to experimental researches for doubly heavy hadrons and make a new test window for NRQCD factorization formalism at various colliders. 

The ultraperipheral heavy ion collisions (UPCs) \cite{Baur:2001jj,Bertulani:2005ru,Baltz:2007kq} is an ideal laboratory to study double heavy hadrons due to its low event multiplicity and efficient signal selection. The equivalent real photon interaction can be studied if ion impact parameter is much larger than the ion radius \footnote{The term ``ultraperipheral" means collisions with distance $b > R_1+R_2$, and should be distinguished from "peripheral" collision where $b\approx R_1+R_2$}, in which the elastic scattering is pure QED process. The unbroken ion will cause less additional calorimetric signals and large signal rapidity gap with produced state, hence led to efficient signal selection. Further more, the photon density is proportional to square of ion charge Z under the equivalent photon approximation (EPA), e.g., the $\gamma-\gamma$ luminosity is enhanced by $82^4$ in Pb-Pb collision, and the production cross sections via photon-photon and photon-gluon collisions for doubly heavy hadrons will be considerable. Therefore, we will explore the doubly heavy baryon and tetraquark production through elastic photon-photon and photon-gluon fusions at the LHC. 

The rest of this paper is organized as follows. In Sect. II, we present the primary formulas employed in the calculation. In Sect. III, the numerical results and discussions toward doubly heavy baryon and tetraquark are performed. The last section is reserved for summary and conclusions.   

\section{FORMULATION} 
The electromagnetic filed of high energy ion can be approximately identified to quasireal photon distribution where the longitudinal part is highly suppressed. Analogous to the Wizsacker–Williams method \cite{vonWeizsacker:1934nji,Williams:1934ad}, the equivalent photon energy spectrum can be formulated by \cite{Cahn:1990jk} 
\begin{align}
	n_{\gamma/A}(\omega) = \dfrac{2 Z^2 \alpha}{\pi}[\xi K_0(\xi)K_1(\xi) - \dfrac{\xi^2}{2}(K_1(\xi)^2-K_0^2(\xi))],
\end{align}
where $\omega$ is photon energy, Z is ion charge, $\alpha$ is the electromagnetic fine structure constant, $\xi = \omega R/\gamma_L\beta$, R is the ionic radius, $\gamma_L,\beta$ are Lorentz factors and $K_{0/1}(\xi)$ is the modified Bessel functions.
\begin{figure}[htbp!]			
	\centering
	\subfigure{\includegraphics[scale=0.28]{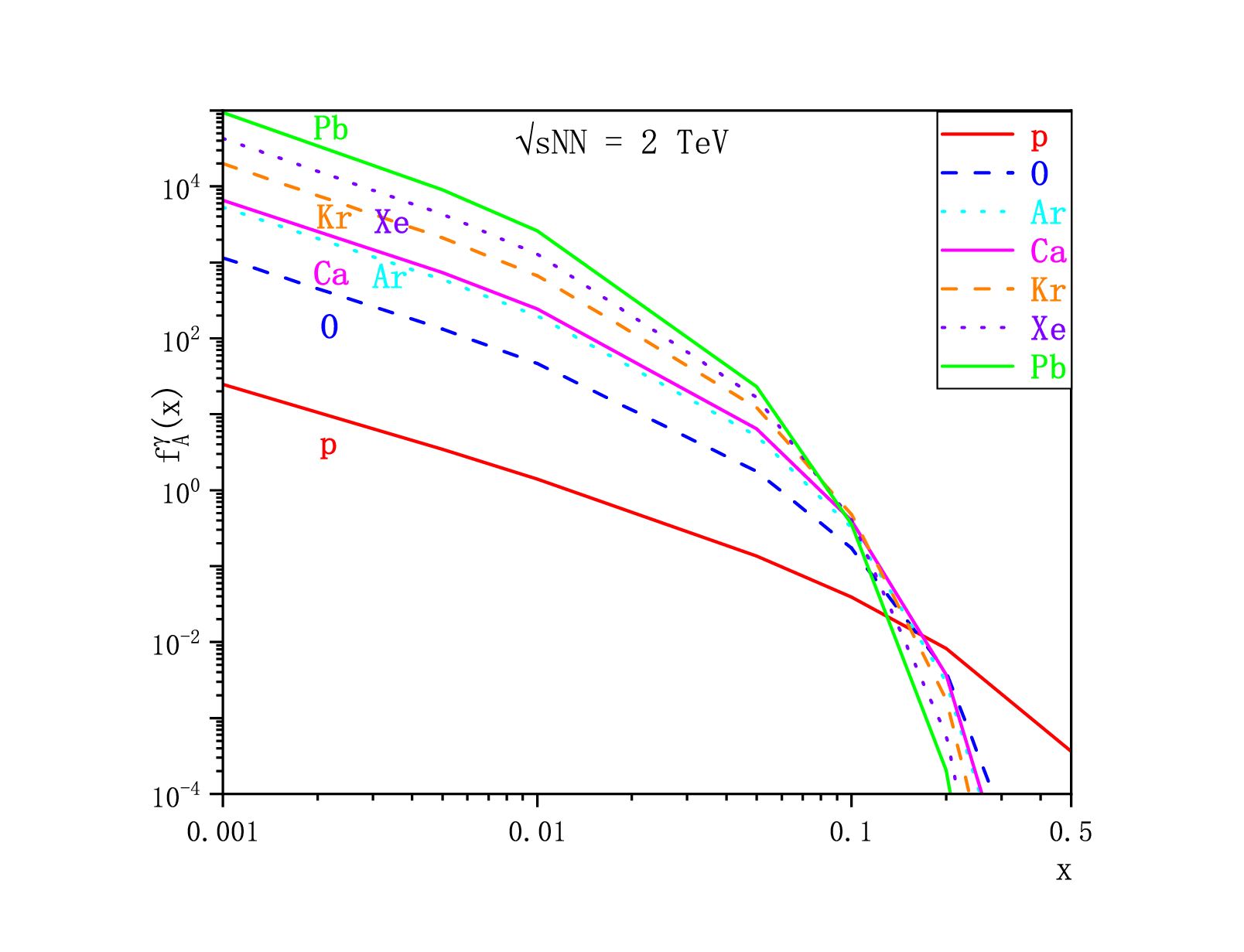}}
	\subfigure{\includegraphics[scale=0.28]{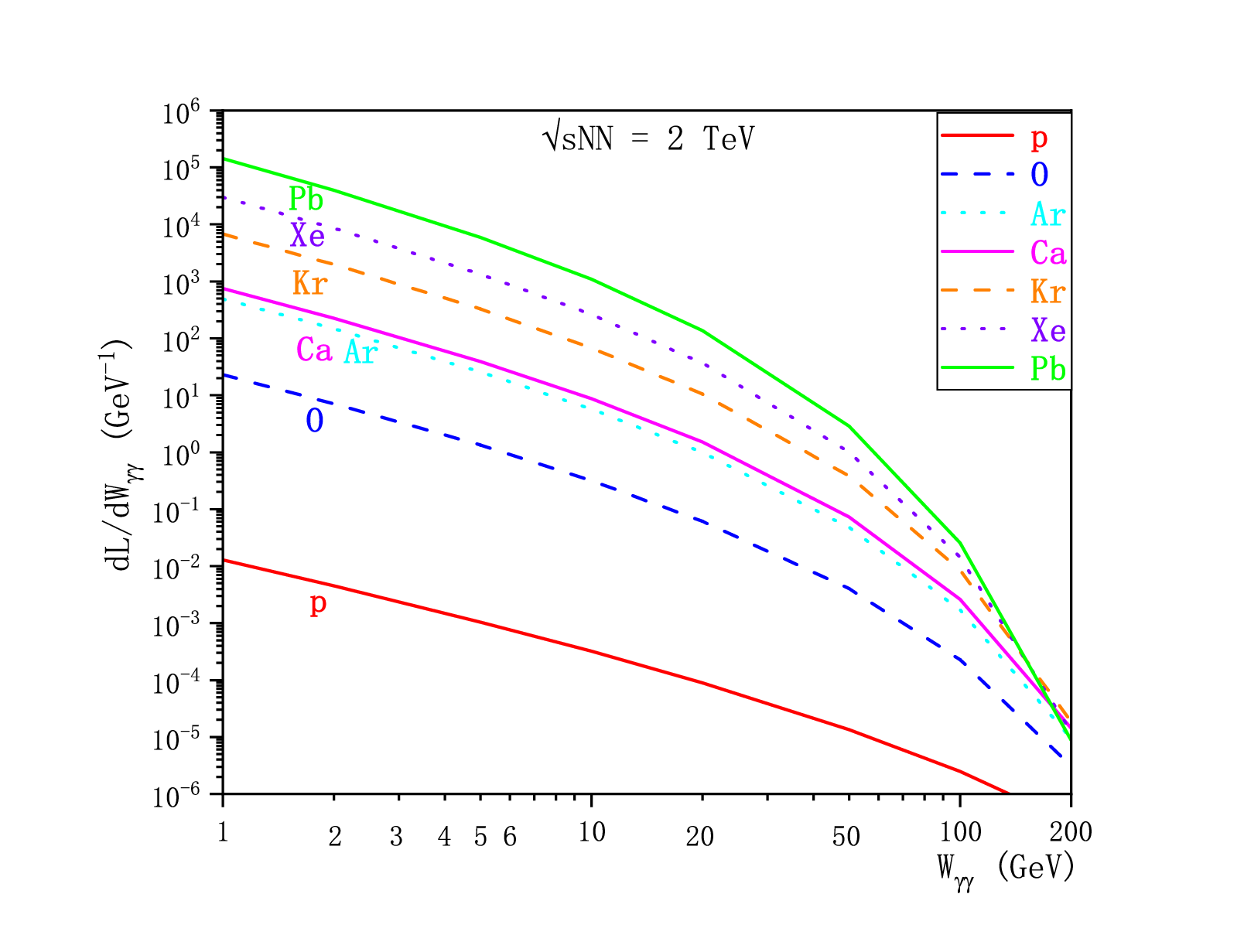}}	
	\caption{Functions of ultraperipheral photon spectra $f_A^{\gamma}(x) = n(\omega)/x$ and comparison of the effective photon-photon luminosities $dL_{\gamma\gamma}/dW_{\gamma\gamma}$ for various ultraperipheral ion collisions with $\sqrt{s_{NN}} = 2\ {\rm TeV}$. }
	\label{FigUPC}
\end{figure}

In this paper, we will refer to elastic photoproduction (photon-photon) cross section for one (two) photon process $A+B\to A+X$ ($A+B\to A+X+B$). The event topologies for those processes are very clean:  very forward ions measured far from the collision point and a few centrally produced particles; the photon momenta can be precisely measured, allowing to reconstruct any missing degrees of freedom in the final state; the background from parton-parton interaction can be sufficiently suppressed if no strongly interaction involved. 

The total cross section for elastic photoproduction of $A+B \to A+X$ can be factorized as
\begin{align}
	\sigma(A+B\to A+X) &= \int \dfrac{d\omega}{\omega}n_{\gamma/A}(\omega) \int dx f_{g/B}(x) \hat{\sigma}(\gamma g\to X),	
\end{align}
where $f_{g/B}(x)$ is the distribution function of gluon for nucleus B. For the elastic photon-photon collision, the total cross section of $A+B \to A+X+B$ is factorized into convolution of photon-photon luminosity and $\gamma\gamma \to X$ cross section,
\begin{align}
	\sigma(A+B\to A+X+B) &= \int \dfrac{d\omega_1}{\omega_1}n_1(\omega_1) \int \dfrac{d\omega_2}{\omega_2}n_2(\omega_2) \hat{\sigma}_{\gamma\gamma\to X}(W_{\gamma\gamma}) \nonumber\\
	       &=\int dW_{\gamma\gamma} \int dY \dfrac{dL_{\gamma\gamma}}{dW_{\gamma\gamma}dY}\hat{\sigma}_{\gamma\gamma\to X}(W_{\gamma\gamma}).
\end{align}
Here $\dfrac{dL_{\gamma\gamma}}{dW_{\gamma\gamma}dY}$ is the equivalent photon luminosity,
\begin{align}
	\dfrac{dL_{\gamma\gamma}}{dW_{\gamma\gamma}dY} = \dfrac{2}{W_{\gamma\gamma}}n_1(\dfrac{W_{\gamma\gamma}}{2}\exp^{Y})n_2(\dfrac{W_{\gamma\gamma}}{2}\exp^{-Y}),
\end{align}
Y is the rapidity of $\gamma-\gamma$ system with $Y = \dfrac{1}{2}\ln \dfrac{\omega_1}{\omega_2}$. The effective photon-photon luminosities $dL_{\gamma}/dW_{\gamma}$ (with Y integrated) for various ion-ion collision are given in FIG. \ref{FigUPC}. Throughout our description, the survival probabilities, which correspond to the probability of scattered ions not to dissociate due to the secondary soft interactions, are set to be 100\%. Detailed calculation of the survival factors may depend on the impact parameters space for each ions. We neglect this probability in the following analysis, the effect is estimated to be less than 20\% as indicated in Ref. \cite{Knapen:2016moh}.
\begin{figure}[htbp!]			
	\centering
	\subfigure{\includegraphics[scale=0.9]{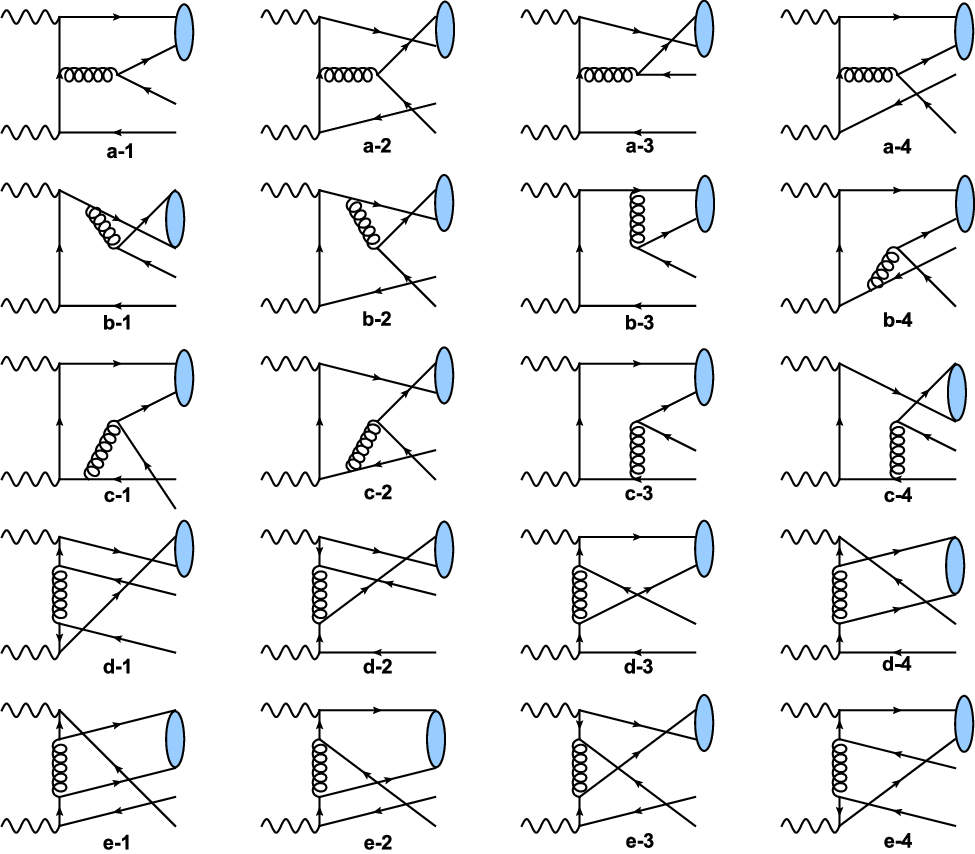}}
	\caption{Half Feynman diagrams for $\gamma\gamma \to (QQ)[n] + \bar{Q}\bar{Q}$, where Q represents the heavy charm or bottom quark, [n] is the spin-color number for the QQ-diquark. The remaining diagrams can be obtained via exchanging initial two photons. The topologies, e.g., $\bm{a\mbox{-}3}$, will not exist for $QQ'[n]$ production, leaving only 20 Feynman diagrams in total.}
	\label{Figrr2Xicc}
\end{figure}

For the elastic photon-photon fusion, there are 40 Feynman diagrams in total for $\gamma\gamma \to (QQ)[n] + \bar{Q} + \bar{Q}$ at leading order (LO)\footnote{We note that there are only 20 feynman diagrams for $\Xi_{bc}$ production.}, half of them are given in Fig.\ref{Figrr2Xicc}, another 20 diagrams can be obtained via exchanging initial photons. The topologies labeled $\bm{b}\text{-}(1,2,3)$ are fragmentation diagrams with one of the final heavy quark fragment into $(QQ)$ diquark. According to NRQCD factorization formalism, the cross section for $\gamma\gamma \to (QQ')[n] + \bar{Q} + \bar{Q'}$ takes the form:
\begin{align}
	\hat{\sigma}_{\gamma\gamma\to H_{QQ'} + \bar{Q} + \bar{Q'}} = \sum_{n} \hat{\sigma}(\gamma\gamma\to (QQ')[n] + \bar{Q} + \bar{Q'})\braket{\mathcal{O}^H(n)},
\end{align}
where $\braket{\mathrm{O}^H(n)}$ is the LDME for $(QQ')[n]$ which represents the inclusive transition probability of $(QQ')[n]$ diquark state into doubly heavy $\Xi_{QQ'}/T_{QQ'}$ hadrons, [n] stands for the spin-color numbers of $(QQ')[n]$ diquark state.   

The short-distance cross section for $\gamma\gamma \to (QQ')[n] + \bar{Q} + \bar{Q'}$,
\begin{align}
	d\hat{\sigma}(\gamma\gamma\to (QQ')[n] + \bar{Q} + \bar{Q'}) = \dfrac{1}{2}\dfrac{1}{2}\frac{1}{2s_{\gamma\gamma}} \overline{\sum}|\mathcal{M}(\gamma\gamma\to (QQ')[n] + \bar{Q} + \bar{Q'})|^2 dPS_3,
	\label{EqSDC}
\end{align}
can be obtained after integrating the phase-space. Here the two $\dfrac{1}{2}$ factors are polarization average for initial photons, $\dfrac{1}{2s_{\gamma\gamma}}$ is the photon-photon flux, an extra $\dfrac{1}{2}$ factor is need if $Q=Q'$ due to identical particles average.

\begin{figure}[htbp!]			
	\centering
	\subfigure{\includegraphics[scale=0.9]{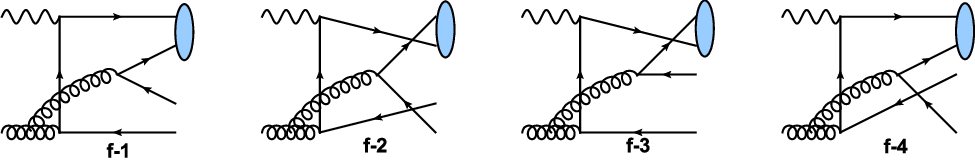}}
	\caption{Extra Feynman diagrams for $\gamma g \to (QQ)[n] + \bar{Q}\bar{Q}$, where Q represents the heavy charm or bottom quark, [n] is the spin-color number for the QQ-diquark. The remaining diagrams can be obtained via exchanging initial photon and gluon.}
	\label{Figrg2Xicc}
\end{figure}

In the elastic photoproduction of $\Xi_{cc}$, apart from similar topologies (one photon is replaced by gluon) in FIG. \ref{Figrr2Xicc}, more topologies are involved, see FIG. \ref{Figrg2Xicc}. An extra 1/8 color factor is needed in Eq. (\ref{EqSDC}).

The hard scattering amplitude $\mathcal{M}$ for $\gamma\gamma (\gamma g)\to (QQ')[n] + \bar{Q} + \bar{Q'}$ can be formulated as:
\begin{align}
	\mathcal{M}((QQ')[n]) = &\epsilon^{\alpha}\epsilon^{\beta} \bar{u}(p_Q,\sigma_1)\gamma_{\alpha_n}s_f(q_{n-1},m_Q)\cdots s_f(q_1,m_Q)\gamma_{\alpha_1} v(p_{\bar{Q}},\sigma_4) \nonumber \\
	& \times \bar{u}(p_{Q'},\sigma_2)\gamma'_{\beta_1}s_f(q'_1,m_{Q'})\cdots s_f(q'_{m-1},m_{Q'})\gamma'_{\beta_m} v(p_{\bar{Q'}},\sigma_3) \nonumber \\
	& \times\mathcal{B}(S,\sigma_1,\sigma_2;p_{QQ'},M_{QQ'})\times \mathcal{C}\times\mathcal{G}.
	\label{Eqamp}
\end{align}
Here, $\epsilon^{\alpha/\beta}$ are the polarization vector for initial states, $\sigma_{i}$ stands for the spin state of final heavy quark, $s_f(q,m)$ is the fermion propagator between two interaction vertexes. $\mathcal{B}(S,\sigma_1,\sigma_2;p_{QQ'},M_{QQ'})$ represents the wave function of heavy diquark $(QQ')[n]$, $\mathcal{C}$ is the SU(3) color factor and $\mathcal{G}$ is the gluon propagator.

In computing the heavy quarkonium production cross sections, the covariant spin-projector method is applied to identify spin-singlet and spin-triplet amplitudes. At the leading order of relative velocity expansion of NRQCD, the standard spin projector \cite{Bodwin:2002cfe} can be written as
\begin{align}
	v(p_{\bar{Q}})\bar{u}(p_{Q'}) = \dfrac{1}{2\sqrt{m_{QQ'}}}\slashed{\epsilon}(p_{QQ'})(p_{QQ'}+m_{QQ'}),
\end{align}
where $\epsilon(p_{QQ'})$ is the polarization vector for spin-triplet state and the projector for spin-singlet state can be obtained by replacing the $\slashed{\epsilon}(p_{QQ'})$ with $\gamma^5$.

To project the amplitude in Eq. (\ref{Eqamp}) into given spin state, we need translate the fermion chain
\begin{align}
	a = \bar{u}(p_Q,\sigma_1)\gamma_{\alpha_n}s_f(q_{n-1},m_Q)\cdots s_f(q_1,m_Q)\gamma_{\alpha_1} v(p_{\bar{Q}},\sigma_4)
\end{align}
into 
\begin{align}
	a &= a^T = v^T(p_{\bar{Q}},\sigma_4) \gamma^T_{\alpha_1} s^T_f(q_1,m_Q) \cdots s^T_f(q_{n-1},m_Q) \gamma^T_{\alpha_n} \bar{u}^T(p_Q,\sigma_1) \nonumber \\
	  &= v^T(p_{\bar{Q}},\sigma_4) \mathcal{C}\mathcal{C}^{-1} \gamma^T_{\alpha_1} \mathcal{C}\mathcal{C}^{-1} s^T_f(q_1,m_Q) \mathcal{C}\mathcal{C}^{-1} \cdots \mathcal{C}\mathcal{C}^{-1} s^T_f(q_{n-1},m_Q) \mathcal{C}\mathcal{C}^{-1} \gamma^T_{\alpha_n} \mathcal{C}\mathcal{C}^{-1} \bar{u}^T(p_Q,\sigma_1) \nonumber \\
	  &= (-1)^{n+1} \bar{u}(p_{\bar{Q}},\sigma_4) \gamma_{\alpha_1} s_f(-q_1,m_Q) \cdots s_f(-q_{n-1},m_Q) \gamma_{\alpha_n} v(p_Q,\sigma_1)
\end{align}
with the help of charge conjugation operator $\mathcal{C} = -\i \gamma^2\gamma^0$ and the relations
\begin{align}
    &\mathcal{C}\mathcal{C}^{-1} = 1, \ \mathcal{C}^{-1}\gamma^T_{\alpha_i}\mathcal{C} = -\gamma_{\alpha_i}, \ 	\mathcal{C}^{-1}s^T_f(q_i,m_Q)\mathcal{C} = s_f(-q_i,m_Q), \ \nonumber \\
    &v^T(p_{\bar{Q}},\sigma_4)\mathcal{C} = -\bar{u}(p_{\bar{Q}},\sigma_4), \
 \mathcal{C}^{-1}\bar{u}^T(p_Q,\sigma_1) = v(p_Q,\sigma_1).
\end{align}

According to SU(3) decomposition: $\bm{3}\bigotimes\bm{3}=\bar{\bm{3}}\bigoplus\bm{6}$, the heavy diquark could be either color anti-triplet $\bar{\bm{3}}$ which is attractive in, e.g., one-gluon exchange potential (OGE) \cite{DeRujula:1975qlm,Jaffe:1999ze}, or color sextuplet $\bm{6}$ which is repulsive. The color factor $\mathcal{C}$ in Eq.(\ref{Eqamp}) is defined as
\begin{align}
	\mathcal{C} = \mathcal{C}_{ijk} = \mathcal{N}_c\times\sum_{m,n}(T^a)_{im}(T^a)_{jn}\times G_{mnk},
\end{align}
where $\mathcal{N}_c = \dfrac{1}{\sqrt{2}}$ is the normalization factor, the factor $G_{mnk}$ stands for the (anti-) symmetric tensor ($\epsilon_{mnk}$) $f_{mnk}$ for the color (anti-triplet $\bar{\bm{3}}$) sextuplet $\bm{6}$ which satisfies the following relations:
\begin{align}
	&\epsilon_{mnk}\epsilon_{m'n'k} = \delta_{mm'}\delta_{nn'} - \delta_{mn'}\delta_{nm'}, \nonumber \\
	&f_{mnk}f_{m'n'k} = \delta_{mm'}\delta_{nn'} + \delta_{mn'}\delta_{nm'}.
\end{align}

In the calculation, the Mathematica package FeynArts \cite{Hahn:2000kx} is used to generate Feynman diagrams; FeynCalc \cite{Mertig:1990an} and FeynCalcFormLink \cite{Feng:2012tk} are used to handle the algebraic calculation; the overall phase space integrals are performed numerically by using the package CUBA \cite{Hahn:2004fe}.

\section{NUMERICAL RESULTS AND DISCUSSIONS}
The LDME for anti-triplet $h_{\bar{\bm{3}}}$ can be related to the matrix element $|\bra{0}\chi^{+}\bm{\sigma}\psi\ket{^3S_1}|^2$, which is the transition of a $Q\bar{Q'}$ pair into a ${^3S_1}$ quarkonium, by assuming that the potentials for binding $Q\bar{Q'}$ and $QQ'$ state are all hydrogen-like. In this way, the radial wave function at the origin $R_{QQ'}(0)$ is related to $h_{\bar{\bm{3}}}$ by:
\begin{align}
	h_{\bar{\bm{3}}} = |\Psi_{QQ'}(0)|^2 = \dfrac{1}{4\pi}|R_{QQ'}(0)|^2.
\end{align}
According to the velocity scaling rule of NRQCD \cite{Bodwin:1994jh}, the LDME for color-sextuplet $h_{\bm{6}}$ is in same order of $h_{\bar{\bm{3}}}$, hence we take $h_{\bm{6}}=h_{\bar{\bm{3}}}$ in our calculation. The input parameters are taken as 
\begin{align}
	\alpha = 1/137.065, m_p = 0.9315\ {\rm GeV},\ m_c = 1.8\ {\rm GeV},\ m_b = 5.1\ {\rm GeV} \nonumber\\
	|\Psi_{cc}(0)|^2 = 0.039\ {\rm GeV^3},\ |\Psi_{bc}(0)|^2 = 0.065\ {\rm GeV^3},\ |\Psi_{bb}(0)|^2 = 0.152\ {\rm GeV^3}, 	   
\end{align}
where the wave functions at the origin are taken from Refs.\cite{Bagan:1994dy,Baranov:1995rc} with above heavy quark masses.

The one-loop formula $\dfrac{\alpha_s(\mu)}{4\pi} = \dfrac{1}{\beta_0 L}$ for the running coupling constant is adopted in our calculation, where $L=\ln(\mu^2/\Lambda_{\rm QCD})^2,\ \beta_0=\frac{11}{3}C_A-\frac{4}{3}T_Fn_f$ with $n_f=4, \Lambda_{\rm QCD} = 297\ {\rm MeV}$ for $\Xi_{cc}$ production, and $n_f=5,\ \Lambda_{\rm QCD} = 214\ {\rm MeV}$ for $\Xi_{bc(bb)}$ production. The renormalization scale is set to be transverse mass of $\Xi_{QQ'}$ with $\mu_r=\sqrt{m^2_{\Xi_{QQ'}}+p_t^2}$. The LHAPDF \cite{Buckley:2014ana} is adopted for the gluon PDFs in ions with data sets labeled ``nCTEQ15WZ" \cite{Kusina:2020lyz}, the factorization scale $\mu_f$ is set to be the same as renormalization scale.   
 
\subsection{Elastic photon-photon production}
The interaction of heavy ions at large impact parameters is purely electromagnetical, such an interaction can be considered as real photon-photon fusion. The generic characteristics of photon-photon fusion in ultraperipheral ion-ion collisions at HL-LHC \cite{Bruce:2018yzs,dEnterria:2022sut} and FCC \cite{Dainese:2016gch,FCC:2018vvp} energies are collected into TABLE. \ref{TabUPC}. Compared to the $e^+e^-$ and p-p collisions, the typical features of photon-photon collision via UPCs are the lack of pileup and highly photon flux boost ($Z^4$). Hence the event signatures will be clear and the reconstruction efficiency can be improved.
\begin{table}[ht]
	\caption{The nucleon-nucleon (NN) c.m. energy $\sqrt{s_{NN}}$, effective charge radius $R_A$ and integrated luminosity per typical run $\mathcal{L}_{int}$ for ultraperipheral collisions at HL-LHC and FCC.}
	\begin{center}
		\scalebox{0.8}{
		\begin{tabular}{|p{2.2cm}<{\centering}|p{1.4cm}<{\centering}|p{1.4cm}<{\centering}|p{1.4cm}<{\centering}|p{1.4cm}<{\centering}|p{1.4cm}<{\centering}|p{1.4cm}<{\centering}|p{1.4cm}<{\centering}|p{1.4cm}<{\centering}||p{1.4cm}<{\centering}|p{1.4cm}<{\centering}|p{1.4cm}<{\centering}|}
			\toprule
			\hline
			System                 & Pb-Pb & Xe-Xe & Kr-Kr & Ar-Ar & Ca-Ca & O-O & p-Pb     & p-p & Pb-Pb & p-Pb     & p-p\\
			\hline
			$\sqrt{s_{NN}}$\ (TeV) & 5.52  & 5.86  & 6.46  & 6.3   & 7.0   & 7.0 & 8.8      & 14  & 39.4  & 62.8     & 100\\
			\hline
			$R_A\ \rm (fm)$        & 7.1   & 6.1   & 5.1   & 4.1   & 4.1   & 3.1 & 0.7,\ 7.1 & 0.7 & 7.1   & 0.7,\ 7.1 & 0.7\\
			\hline
			$\mathcal{L}_{int}$    & $5\ \rm nb^{-1}$ & $30\ \rm nb^{-1}$ & $120\ \rm nb^{-1}$ & $1.1\ \rm pb^{-1}$ & $0.8\ \rm pb^{-1}$ & $12\ \rm pb^{-1}$ & $1\ \rm pb^{-1}$ & $150\ \rm fb^{-1}$ & $110\ \rm nb^{-1}$ & $29\ \rm pb^{-1}$ & $1\ \rm ab^{-1}$\\
			\hline   		
		\end{tabular}
	    }
	\end{center}    
	\label{TabUPC}
\end{table}

\begin{figure}[htbp!]			
	\centering
	\subfigure{\includegraphics[scale=0.3]{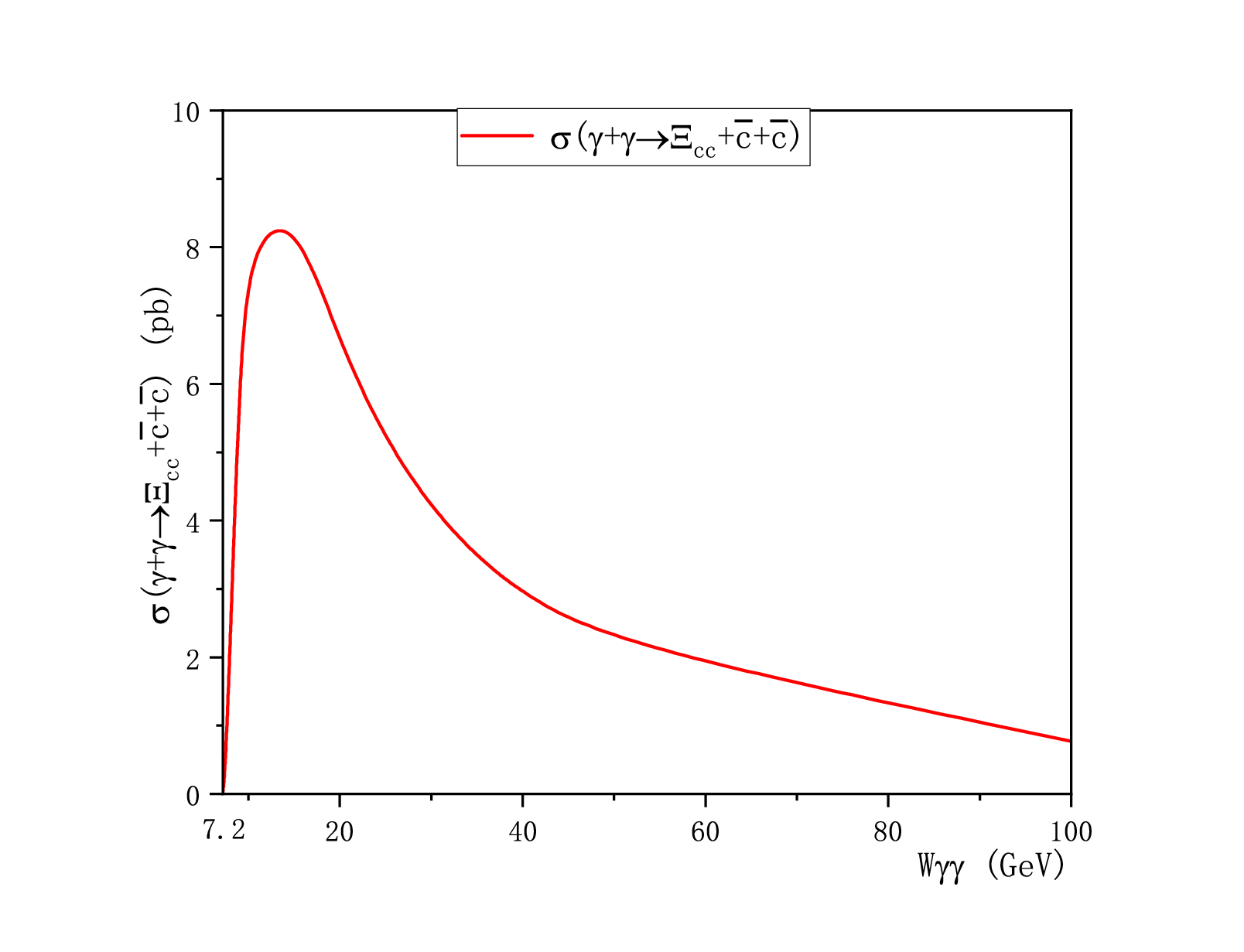}}
	\caption{The cross section for $\gamma+\gamma\to \Xi_{cc}[{^3S_1}\mbox{-}\bar{\bm{3}}+{^1S_0}\mbox{-}\bm{6}] +\bar{c}\bar{c}$ versus c.m. energy ($W_{\gamma\gamma}$) of photon-photon fusion.}
	\label{FigXiccW}
\end{figure}

The photon-photon cross section for $\Xi_{cc}$ versus $W_{\gamma\gamma}$ within default parameters is given in FIG. \ref{FigXiccW}, the contributions from color anti-triplet $\bar{\bm{3}}$ and sextuplet $\bm{6}$ are collected. The cross section reaches its maximal value just several GeV above the threshold, and decreases with c.m. energy of photon-photon system. As the effective photon-photon luminosities decrease dramatically with $W_{\gamma\gamma}$, the main contribution can be only related to small $W_{\gamma\gamma}$ values, typically from $4m_c$ to 50 GeV. The cross sections for each spin-color state of $\Xi_{cc}$ are listed in TABLE. \ref{Tabrr2Xicc}. The contribution from $cc[{^1S_0}\mbox{-}\bm{6}]$ is only $3\sim5\%$ to that of $cc[{^3S_1}\mbox{-}\bar{\bm{3}}]$, the ratio holds also for $\Xi_{bb}$.

\begin{table}[htbp!]
	\caption{The cross sections for $\gamma+\gamma\to \Xi_{cc}[cc,n]+\bar{c}\bar{c}$ through UPCs at the HL-LHC and FCC.}
	\begin{center}
		\scalebox{0.8}{
		\begin{tabular}{|p{2.cm}<{\centering}|p{2.2cm}<{\centering}|p{2.2cm}<{\centering}|p{2.2cm}<{\centering}|p{2.cm}<{\centering}|p{2.cm}<{\centering}|}
			\toprule
			\hline
			Collisions & $\sqrt{s_{NN}}$\ (TeV) & $ \Xi_{cc}[cc,{^3S_1}\mbox{-}\bar{\bm{3}}] $ & $\Xi_{cc}[cc,{^1S_0}\mbox{-}\bm{6}]$ & Total & $N_{\Xi_{cc}}$ \\
			\hline
			Pb-Pb      & 5.52  & 270 nb    & 9.53 nb      & 279.5 nb  &$1.40\times10^3$\\
			\hline
			Xe-Xe      & 5.86  & 65.9 nb   & 2.38 nb      & 68.28 nb  &$2.05\times10^3$\\
			\hline
			Kr-Kr      & 6.46  & 17.8 nb    & 0.663 nb    & 18.46 nb  &$2.21\times10^3$\\
			\hline
			Ar-Ar      & 6.3   & 1.36 nb    & 0.0518 nb   & 1.411 nb  &$1.55\times10^3$\\
			\hline
			Ca-Ca      & 7.0   & 2.31 nb    & 0.0886 nb   & 2.398 nb  &$1.92\times10^3$\\
			\hline
			O-O        & 7.0   & 77.1 pb     & 3.03 pb    & 80.13 pb  &$9.61\times10^2$\\
			\hline
			p-Pb       & 8.8   & 203 pb     & 8.33 pb     & 211.33 pb &$2.11\times10^2$\\
			\hline
			p-p        & 14    & 89.6 fb     & 3.99 fb    & 93.59 fb  &$1.43\times10^4$\\
			\hline
			\hline
			Pb-Pb      & 39.4  & 1780 nb    & 74.5 nb     & 1854 nb   &$2.04\times10^5$\\
			\hline
			p-Pb       & 62.8  & 728 pb    & 32.8 pb      & 760.8 pb  &$2.20\times10^4$\\
			\hline
			p-p        & 100   & 233 fb     & 10.9 fb     & 243.9 fb  &$2.44\times10^5$\\
			\hline		
		\end{tabular}
	    }
	\end{center}
	\label{Tabrr2Xicc}
\end{table}

Supposing the integrated luminosities in TABLE \ref{TabUPC} and aggregating the contributions for diquark in all spin-color structures, the produced $\Xi_{cc}$ numbers via various UPCs at the HL-LHC are around $10^{3}$. As the collision energies and luminosities are highly improved at the FCC, the yields for $\Xi_{cc}$ can be increased by one or two magnitudes, reaching $2.44\times10^{5}$ for p-p in 100 TeV with a integrated luminosity of $1\ {\rm ab^{-1}}$. To estimate the event, we set the relative possibilities for various light quarks as $u:d:s \sim 1:1:0.3$, and the reconstruction channel $Br(\Xi_{cc}^{++}\to\Lambda_c^+ K^-\pi^+\pi^+) \approx 10\%$ \cite{Yu:2017zst}, $Br(\Lambda_c^+\to pK^+\pi^+)\approx 5\%$ \cite{LHCb:2013hvt}. Considering some detection efficiencies, $\Xi_{cc}^{++}$ events via ultraperipheral collisions may be expected at the future FCC.
\begin{figure}[htbp!]			
	\centering
	\subfigure{\includegraphics[scale=0.28]{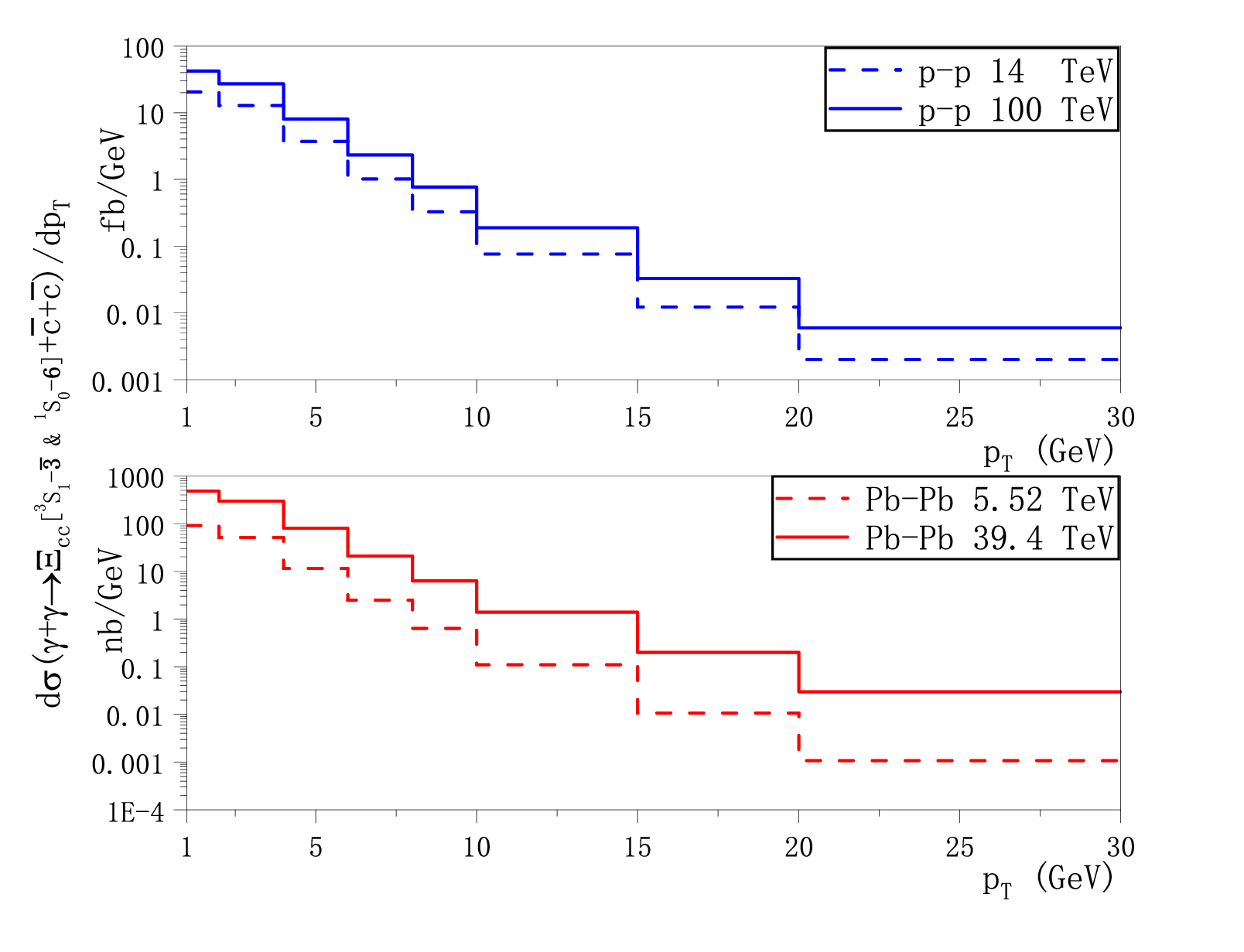}}
	\subfigure{\includegraphics[scale=0.28]{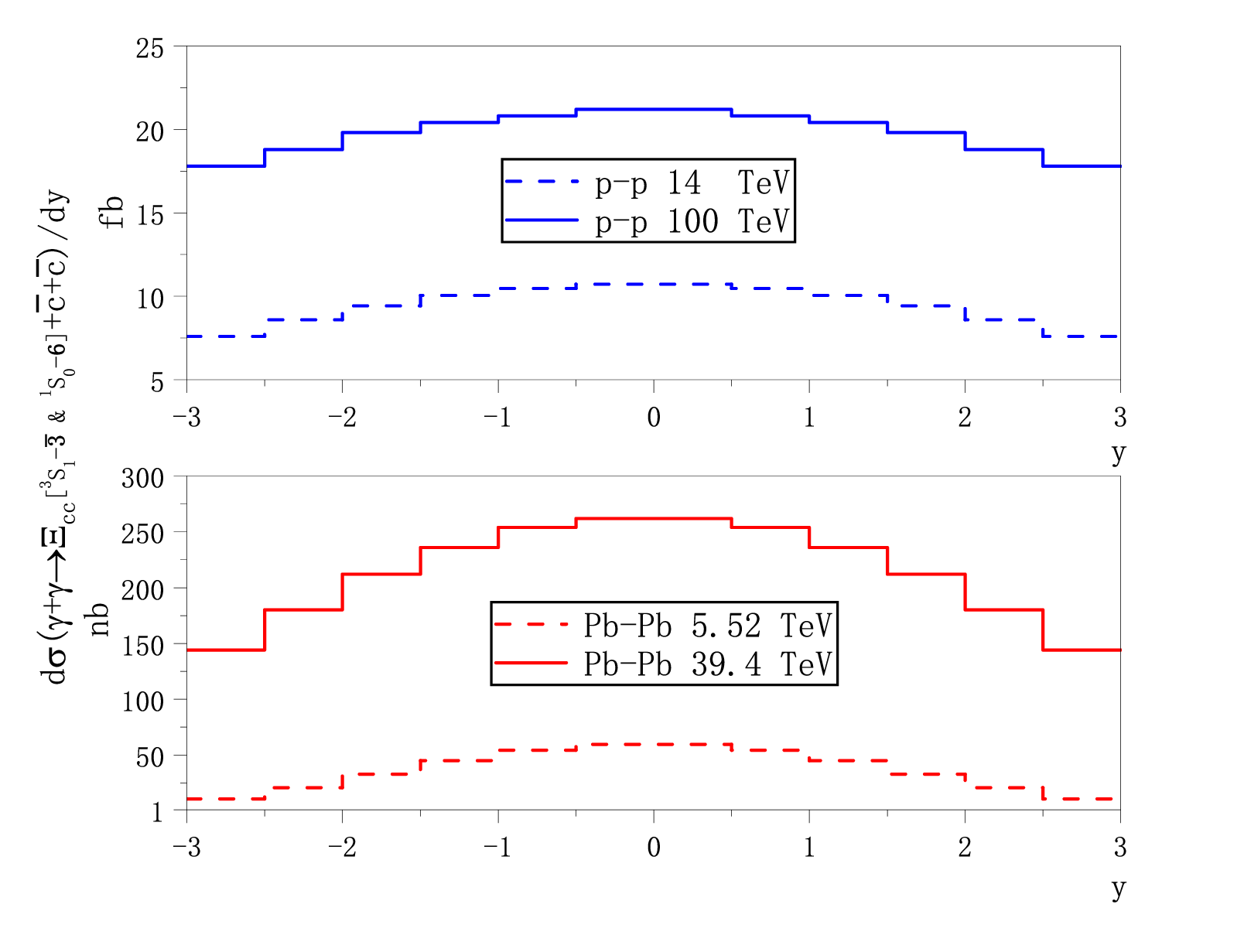}}
	\caption{The transverse momentum $p_T$ and rapidity y distributions for $\Xi_{cc}$ production via ultraperipheral collisions. Here, for the $p_T$ distribution, y is cut to be $[-3,3]$; for the y distribution, $p_T$ is cut to be $1\mbox{-}30$ GeV.}
	\label{Figrr2Xiccpty}
\end{figure}

\begin{table}[ht]
	\caption{The cross sections for $\Xi_{cc}[cc,{^3S_1}\mbox{-}\bar{\bm{3}}]$ ($\Xi_{cc}[cc,{^1S_0}\mbox{-}\bm{6}]$) (in unit of nb) under different $m_c$ and renormalization scales through ultraperipheral Pb-Pb collision at 5.52 TeV.}
	\begin{center}
		\scalebox{0.9}{
			\begin{tabular}{|l|c|c|c|}
				\toprule
				\hline
				\diagbox{$\mu$}{$m_c$ (GeV)} & \hspace{1.cm}1.7\hspace{1.cm} & \hspace{1.cm}1.8\hspace{1.cm} & \hspace{1.cm}1.9\hspace{1.cm} \\
				\hline
				$\dfrac{1}{2}\sqrt{4m_c^2+p_T^2}$ & 752 (25.8) & 496 (16.9) & 334 (11.4) \\
				\hline
				$\sqrt{4m_c^2+p_T^2}$             & 404 (14.3) & 271 (9.53) & 185 (6.48) \\
				\hline
				$2\sqrt{4m_c^2+p_T^2}$            & 252 (9.09) & 170 (6.10) & 117 (4.18) \\
				\hline		
			\end{tabular}
		}
	\end{center}
	\label{TabUncertainty}
\end{table}
As the number of events corresponding to $\gamma+\gamma\to \Xi_{cc}[cc,{^3S_1}\mbox{-}\bar{\bm{3}} + {^1S_0}\mbox{-}\bm{6}] + \bar{c}\bar{c}$ is considerable, it is worthy to perform a more elaborate phenomenological analysis. The transverse momentum and rapidity distributions of $\Xi_{cc}$ through UPCs are given in FIG. \ref{Figrr2Xiccpty}, with the $p_T = 1-30$ GeV and $y=[-3,3]$. The cross sections decease rapidly versus high $p_T$, showing a logarithmic dependence of $p_T$. As the UPCs are characterized by a large rapidity gap between the produced state and the interacting nucleus accompanied by forward neutron emission from the de-excitation of nucleus \cite{Baltz:2007kq}, the resulting rapidity distribution is relative narrow and centered at midrapidity. For the equal energy beam collisions, the rapidity distributions show a symmetric behavior. To estimate the theoretical uncertainties caused charm mass and renormalization scale, we set the charm mass to be 1.7, 1.8 and 1.9 GeV, and the renormalization scales are choosed to be $\dfrac{1}{2}$, 1 and 2 times of  $\Xi_{cc}$ transverse mass, see TABLE \ref{TabUncertainty}. 

\begin{table}[ht]
	\caption{The cross sections for $\gamma+\gamma \to \Xi_{bc}[bc,n]+\bar{b}\bar{c}$ through UPCs at the HL-LHC and FCC.}
	\begin{center}
		\scalebox{0.8}{
		\begin{tabular}{|p{2.cm}<{\centering}|p{2.2cm}<{\centering}|p{2.2cm}<{\centering}|p{2.2cm}<{\centering}|p{2.2cm}<{\centering}|p{2.2cm}<{\centering}|p{2.2cm}<{\centering}|}
			\toprule
			\hline
			Collisions & $\sqrt{s_{NN}}$\ (TeV) & $\Xi_{bc}[bc,{^3S_1}\mbox{-}\bar{\bm{3}}]$ & $\Xi_{bc}[bc,{^3S_1}\mbox{-}\bm{6}]$ &$\Xi_{bc}[bc,{^1S_0}\mbox{-}\bar{\bm{3}}]$ & $\Xi_{bc}[bc,{^1S_0}\mbox{-}\bm{6}]$ & Total \\
			\hline
			Pb-Pb      & 5.52  & 850 pb   & 425 pb   & 335 pb   & 167.5 pb & 1777 pb  \\
			\hline
			Xe-Xe      & 5.86  & 221 pb   & 110 pb   & 88 pb    & 44 pb    & 463 pb   \\
			\hline
			Kr-Kr      & 6.46  & 64.3 pb  & 32.1 pb  & 25.8 pb  & 12.9 pb  & 135.1 pb \\
			\hline
			Ar-Ar      & 6.3   & 5.16 pb  & 2.58 pb  & 2.08 pb  & 1.04 pb  & 10.86 pb \\
			\hline
			Ca-Ca      & 7.0   & 8.94 pb  & 4.47 pb  & 3.62 pb  & 1.81 pb  & 18.84 pb \\
			\hline
			O-O        & 7.0   & 315 fb   & 157 fb   & 128 fb   & 64 fb    & 664 fb   \\
			\hline
			p-Pb       & 8.8   & 910 fb   & 455 fb   & 374 fb   & 187 fb   & 1926 fb  \\
			\hline
			p-p        & 14    & 0.472 fb & 0.236 fb & 0.197 fb & 0.098 fb & 1.003 fb \\
			\hline
			\hline
			Pb-Pb      & 39.4  & 8.32 nb  & 4.16 nb  & 3.44 nb  & 1.72 nb  & 17.64 nb \\
			\hline
			p-Pb       & 62.8  & 3.92 pb  & 1.96 pb  & 1.65 pb  & 0.825 pb & 8.355 pb \\
			\hline
			p-p        & 100   & 1.36 fb  & 0.68 fb  & 0.58 fb  & 0.29 fb  & 2.91 fb  \\
			\hline		
		\end{tabular}
	    }
	\end{center}
	\label{Tabrr2Xibc}
\end{table}

\begin{table}[ht]
	\caption{The cross sections for $\gamma+\gamma \to \Xi_{bb}[bb,n]+\bar{b}\bar{b}$ through UPCs at the LHC and FCC.}
	\begin{center}
		\scalebox{0.8}{
		\begin{tabular}{|p{2.cm}<{\centering}|p{2.2cm}<{\centering}|p{4.5cm}<{\centering}|p{4.5cm}<{\centering}|p{2.cm}<{\centering}|}
			\toprule
			\hline
			Collisions & $\sqrt{s_{NN}}$\ (TeV) & $\sigma(\gamma\gamma\to \Xi_{bb}[bb,{^3S_1}\mbox{-}\bar{\bm{3}}]+\bar{b}\bar{b})$ & $\sigma(\gamma\gamma\to \Xi_{bb}[bb,{^1S_0}\mbox{-}\bm{6}]+\bar{b}\bar{b})$ & Total \\
			\hline
			Pb-Pb      & 5.52  & 37.0 pb   & 1.12 pb   & 38.12 pb  \\
			\hline
			Xe-Xe      & 5.86  & 10.2 pb   & 0.323 pb  & 10.52 pb  \\
			\hline
			Kr-Kr      & 6.46  & 3.18 pb   & 0.105 pb  & 3.285 pb  \\
			\hline
			Ar-Ar      & 6.3   & 265 fb    & 9.09 fb   & 274.1 fb  \\
			\hline
			Ca-Ca      & 7.0   & 470 fb    & 16.3 fb   & 486.3 fb  \\
			\hline
			O-O        & 7.0   & 17.4 fb   & 0.629 fb  & 18.02 fb  \\
			\hline
			p-Pb       & 8.8   & 54.1 fb   & 2.08 fb   & 56.18 fb  \\
			\hline
			p-p        & 14    & 32 ab     & 1.4 ab    & 33.4 ab   \\
			\hline
			\hline
			Pb-Pb      & 39.4  & 514 pb    & 20.4 pb   & 534.4 nb  \\
			\hline
			p-Pb       & 62.8  & 273 fb    & 12 fb     & 285 fb    \\
			\hline
			p-p        & 100   & 101 ab    & 4.74 ab   & 105.7 ab  \\
			\hline		
		\end{tabular}
	    }
	\end{center}
	\label{Tabrr2Xibb}
\end{table}
We also study the production cross sections for $\Xi_{bc/bb}$ via various UPCs at HL-LHC and FCC, the numerical results for each spin-color states are listed in TABLE \ref{Tabrr2Xibc} and \ref{Tabrr2Xibb}. As the heavy constituent for $bc$-quark are different, the exchange asymmetry for identical particle is not hold, hence all diquark structures for spin-color states in $bc[{^3S_1}\mbox{-}\bar{\bm{3}}],\ bc[{^3S_1}\mbox{-}\bm{6}],\ bc[{^1S_0}\mbox{-}\bar{\bm{3}}]$ and $bc[{^1S_0}\mbox{-}\bm{6}]$ will contribute in a comparable level. Furthermore, the cross section for color anti-triplet $\bar{\bm{3}}$ is two times of sextuplet $\bm{6}$ in the same spin state (${^1S_0}$ or ${^3S_1}$) due to SU(3) algebra. At the HL-LHC, only tens of $\Xi_{bc}$ could be produced, and in thousand for FCC. For the $\Xi_{bb}$, the production cross sections are further suppressed. Hence the phenomenological investigation for $\Xi_{bc/bb}$ through ultraperipheral ion-ion collisions may not be feasible.

\subsection{Elastic photoproduction}
In this subsection, we shall discuss the cross sections for the $\Xi_{cc(bc/bb)}$ production through elastic photoproduction $A+B \to A + \Xi_{cc(bc/bb)} + \bar{c}\bar{c}(\bar{b}\bar{c}/\bar{b}\bar{b}) +X$ at the LHC, the productions for doubly heavy tetraquark are also discussed. Different from photoproduction in the electron-ion collision, where only $\gamma+g$ channel will contribute with one photon produced by electron and one gluon produced by ion, $g+\gamma$ channel is also need to be take into account. For the same ion collision, e.g., Pb-Pb, the two channels will led to equal cross section; while for the p-Pb collision, the luminosity is enhanced by $1\times {\rm Z}^2$ (Z is charge number of Pb) for $g+\gamma$ channel, and only enhanced by $1\times {\rm A}$ (A is nucleus number of Pb) for $\gamma+g$ channel, thus the contribution from $\gamma+g$ is negligible. In the following analysis, the contributions for both channels are given. 
\begin{table}[htbp!]
	\caption{The cross sections for $g+\gamma\to \Xi_{cc}[cc,n]+\bar{c}\bar{c}$ through elastic photoproduction at the HL-LHC and FCC. The cross sections in brackets are the contributions from $\gamma+g$ channel, which are different from $g+\gamma$ channel in the p-Pb collision; while the two contributions are absolutely equal for same ions collision. The total cross sections contains all the $g+\gamma$ and $\gamma+g$ channels.}
	\begin{center}
		\scalebox{0.8}{
			\begin{tabular}{|p{2.cm}<{\centering}|p{2.2cm}<{\centering}|p{2.8cm}<{\centering}|p{2.8cm}<{\centering}|p{2.cm}<{\centering}|p{2.cm}<{\centering}|}
				\toprule
				\hline
				Collisions & $\sqrt{s_{NN}}$\ (TeV) & $ \Xi_{cc}[cc,{^3S_1}\mbox{-}\bar{\bm{3}}] $ & $\Xi_{cc}[cc,{^1S_0}\mbox{-}\bm{6}]$ & Total & $N_{\Xi_{cc}}$ \\
				\hline
				Pb-Pb      & 5.52  & 82.9 $\mu$b         & 7.09 $\mu$b          & 179.98 $\mu$b &$9.00\times10^5$\\
				\hline
				Xe-Xe      & 5.86  & 24.9 $\mu$b         & 2.14 $\mu$b          & 54.08 $\mu$b  &$1.62\times10^6$\\
				\hline 
				Kr-Kr      & 6.46  & 8.34 $\mu$b         & 0.72 $\mu$b          & 18.12 $\mu$b  &$2.17\times10^6$\\
				\hline
				Ar-Ar      & 6.3   & 1.08 $\mu$b         & 0.094 $\mu$b         & 2.35 $\mu$b   &$2.58\times10^6$\\
				\hline
				Ca-Ca      & 7.0   & 1.46 $\mu$b         & 0.127 $\mu$b         & 3.17 $\mu$b   &$2.53\times10^6$\\
				\hline
				O-O        & 7.0   & 108 nb              & 9.44 nb              & 234.88 nb     &$2.81\times10^6$\\
				\hline
				p-Pb       & 8.8   & 628 (44.2) nb       & 54.2 (4.01) nb       & 730.41 nb     &$7.30\times10^5$\\
				\hline
				p-p        & 14    & 325 pb              & 29.6 pb              & 709.2 pb      &$1.06\times10^8$\\
				\hline
				\hline
				Pb-Pb      & 39.4  & 374 $\mu$b          & 34.3 $\mu$b          & 816.6 $\mu$b  &$8.98\times10^7$\\
				\hline
				p-Pb       & 62.8  & 2.75 (0.14) $\mu$b  & 0.25 (0.013) $\mu$b  & 3.15  $\mu$b  &$9.13\times10^7$\\
				\hline
				p-p        & 100   & 1090 pb             & 103 pb               & 2386 pb       &$2.38\times10^9$\\
				\hline		
			\end{tabular}
		}
	\end{center}
	\label{Tabgr2Xicc}
\end{table}

\begin{figure}[htbp!]			
	\centering
	\subfigure{\includegraphics[scale=0.28]{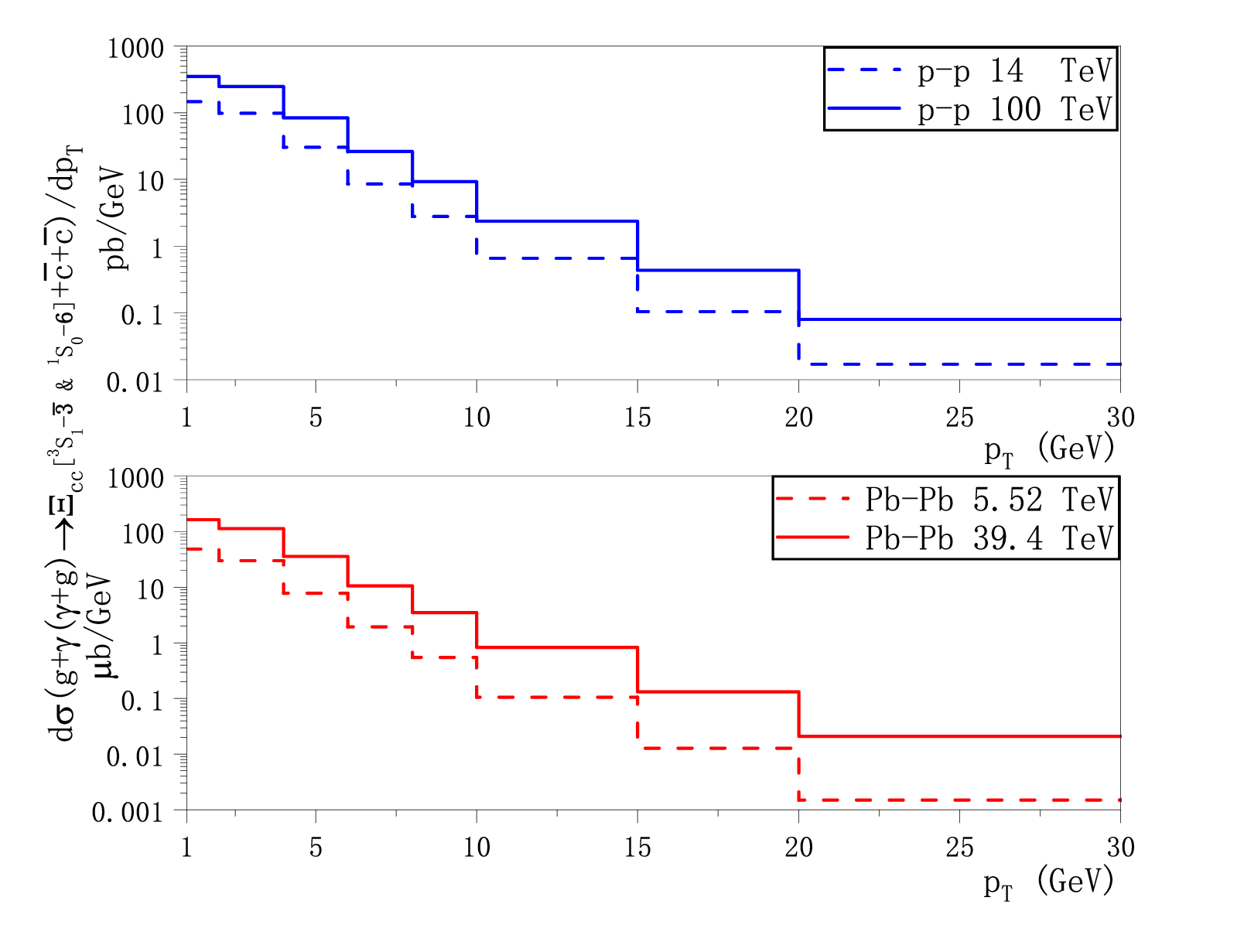}}
	\subfigure{\includegraphics[scale=0.28]{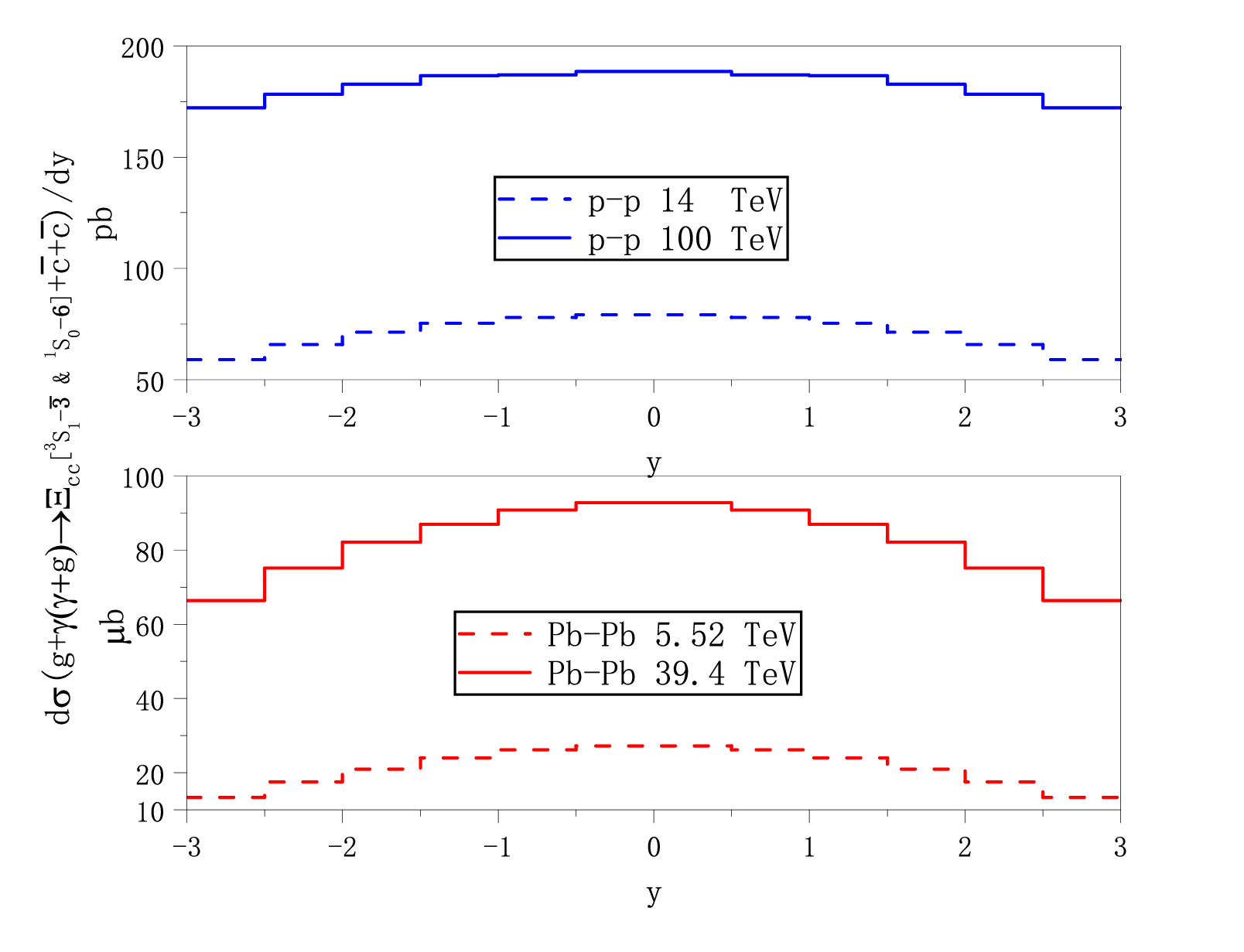}}
	\caption{The transverse momentum $p_T$ and rapidity y distributions for $\Xi_{cc}$ production via semielastic ion-ion collisions. Here, for the $p_T$ distribution, y is cut to be $[-3,3]$; for the y distribution, $p_T$ is cut to be $1\mbox{-}30$ GeV.}
	\label{Figgr2Xiccpty}
\end{figure}

The elastic photoproduction cross sections for each spin-color state of $\Xi_{cc}$ are listed in TABLE. \ref{Tabgr2Xicc}. Compared with elastic photon-photon production, the production ratios for $[cc,{^1S_0}\mbox{-}\bm{6}]/[cc,{^3S_1}\mbox{-}\bar{\bm{3}}]$ increase to $8\sim9\%$. Supposing the integrated luminosities in TABLE \ref{TabUPC} and collecting all the diquark structures, the produced $\Xi_{cc}$ numbers via various elastic photoproduction at the HL-LHC and FCC are estimated in TABLE \ref{Tabgr2Xicc}. In Pb-Pb collision with $\sqrt{s_{NN}} = 5.52$ TeV, the produced $\Xi_{cc}$ is around $9\times10^{5}$, and increases to $10^{8}$ for 14 TeV p-p collision due its high luminosity. To estimate the event, we adopt the reconstruction channel $Br(\Xi_{cc}^{++}\to\Lambda_c^+ K^-\pi^+\pi^+) \approx 10\%$ \cite{Yu:2017zst}, $Br(\Lambda_c^+\to pK^+\pi^+)\approx 5\%$ \cite{LHCb:2013hvt}. The event numbers for $\Xi_{cc}^{++}$ in this reconstruction channel are $1.95\times10^{3}$ for Pb-Pb collision, $2.3\times10^{5}$ for p-p collision. As the collision energies and luminosities are highly improved at the FCC, the yields for $\Xi_{cc}$ can be increased by one or two magnitudes, leaving the possibility for further phenomenological investigation toward $\Xi_{cc}$. As the physical potential for observing doubly charmed baryon via $g+\gamma (\gamma+g)\to \Xi_{cc}[cc, {^3S_1}\mbox{-}\bar{\bm{3}} + {^1S_0}\mbox{-}\bm{6}] + \bar{c}\bar{c}$ channel is large, we perform a detailed transverse momentum and rapidity distributions of $\Xi_{cc}$ in FIG. \ref{Figgr2Xiccpty}, with the $p_T$ is cut to be $1\mbox{-}30$ GeV and y is set to be $[-3,3]$.

Considering the number of produced $[cc]$-diquark is large, it is also possible to form a compact heavy-heavy tetraquark $T_{cc}$ composed of cc-diquark and light-antidiquark (e.g., $\bar{u}\bar{d}$). The production mechanism of $T_{cc}$ is similar to $\Xi_{cc}$: I. the generation of $[cc,\ n]$ diquark at short distance; II. the subsequent formation of $T_{cc}$ via combining two light quark at long distance. The heavy-heavy diquark cluster in color anti-triplet may served as heavy antiquark due to diquark-antiquark symmetry. Therefore, in the heavy quark limit, its fragmentation probability for combining light freedom to form doubly heavy tetraquark can be approximately described as the probability to form heavy baryon from a heavy quark. The combining probability is described by the fragmentation function $D_{\Lambda_c/c}(z)$ of charm quark to charm baryon at the heavy quark limit. The fragmentation fraction for $c\to \Lambda_c^+$ is measured to be $20.4\%$ \cite{ALICE:2021dhb} at the LHC. In this way, the production cross sections for $T_{cc}^+$ can be estimated by $\sigma(\Xi_{cc})\times 20.4\%$, yield a total cross sections of 36.71 $\mu b$ for Pb-Pb collision at 5.52 TeV, 14.46 pb for p-p collision at 14 TeV. Thus the produced $T_{cc}^{+}$ numbers could be $1.84\times10^{5}$ and $2.16\times10^{7}$ respectively. Considering $T_{cc}^{+}$ is reconstructed by $D^0D^0\pi^+$ channel (supposed to be $100\%$), with subsequent $D^0$ decay $Br(D^0 \to K^-\pi^+) = 3.94\%$ \cite{ParticleDataGroup:2022pth}, hence the event number for $T_{cc}^{+}$ could be 285 and $3.53\times10^{4}$ via elastic photoproduction of Pb-Pb and p-p collision. As for the FCC, the events could be further extended due to its better performance.    

\begin{table}[ht]
	\caption{The cross sections for $g+\gamma\to \Xi_{bc}[bc,n]+\bar{b}\bar{c}$ through elastic photoproduction at the HL-LHC and FCC. The cross sections in brackets are the contributions from $\gamma+g$ channel, which are different from $g+\gamma$ channel in the p-Pb collision; while the two contributions are absolutely equal for same ions collision. The total cross sections contains all the $g+\gamma$ and $\gamma+g$ channels.}
	\begin{center}
		\scalebox{0.8}{
			\begin{tabular}{|p{2.cm}<{\centering}|p{2.2cm}<{\centering}|p{3.0cm}<{\centering}|p{3.0cm}<{\centering}|p{3.0cm}<{\centering}|p{3.0cm}<{\centering}|p{2.2cm}<{\centering}|}
				\toprule
				\hline
				Collisions & $\sqrt{s_{NN}}$\ (TeV) & $\Xi_{bc}[bc,{^3S_1}\mbox{-}\bar{\bm{3}}]$ & $\Xi_{bc}[bc,{^3S_1}\mbox{-}\bm{6}]$ &$\Xi_{bc}[bc,{^1S_0}\mbox{-}\bar{\bm{3}}]$ & $\Xi_{bc}[bc,{^1S_0}\mbox{-}\bm{6}]$ & Total \\
				\hline
				Pb-Pb      & 5.52  & 619 nb   & 530 nb   & 184 nb   & 129 nb & 2924 nb  \\
				\hline
				Xe-Xe      & 5.86  & 193 nb   & 165 nb   & 57.9 nb    & 40.3 nb    & 912.4 nb   \\
				\hline
				Kr-Kr      & 6.46  & 67.8 nb  & 57.8 nb  & 20.4 nb  & 14.1 nb  & 320.2 nb \\
				\hline
				Ar-Ar      & 6.3   & 8.88 nb  & 7.57 nb  & 2.67 nb  & 1.85 nb  & 41.94 nb \\
				\hline
				Ca-Ca      & 7.0   & 12.3 nb  & 10.5 nb  & 3.72 nb  & 2.57 nb  & 58.18 nb \\
				\hline
				O-O        & 7.0   & 932 pb   & 792 fb   & 281 pb   & 194 pb    & 4398 pb   \\
				\hline
				p-Pb       & 8.8   & 5.07 (0.48) nb   & 4.32 (0.40) nb   & 1.52 (0.15) nb   & 1.06 (0.10) nb   & 13.1 nb  \\
				\hline
				p-p        & 14    & 3.64 pb & 3.06 pb & 1.12 pb & 0.76 pb   & 17.16 pb \\
				\hline
				\hline
				Pb-Pb      & 39.4  & 4334 nb  & 3635 nb  & 1338 nb  & 907 nb  & 20.4 $\mu$b \\
				\hline
				p-Pb       & 62.8  & 32.5 (2.10) nb  & 27.2 (1.75) nb  & 10.0 (0.65) nb  & 6.81 (0.44) nb & 81.45 nb \\
				\hline
				p-p        & 100   & 15.7 pb  & 13.0 pb  & 4.91 pb  & 3.29 pb  & 73.8 pb  \\
				\hline		
			\end{tabular}
		}
	\end{center}
	\label{Tabgr2Xibc}
\end{table}

\begin{table}[ht]
	\caption{The cross sections for $g+\gamma \to \Xi_{bb}[bb,n]+\bar{b}\bar{b}$ through elastic photoproduction at the HL-LHC and FCC. The cross sections in brackets are the contributions from $\gamma+g$ channel, which are different from $g+\gamma$ channel in the p-Pb collision; while the two contributions are absolutely equal for same ions collision. The total cross sections contains all the $g+\gamma$ and $\gamma+g$ channels.}
	\begin{center}
		\scalebox{0.8}{
			\begin{tabular}{|p{2.cm}<{\centering}|p{2.2cm}<{\centering}|p{4.8cm}<{\centering}|p{4.8cm}<{\centering}|p{2.cm}<{\centering}|}
				\toprule
				\hline
				Collisions & $\sqrt{s_{NN}}$\ (TeV) & $\sigma(g\gamma\to \Xi_{bb}[bb,{^3S_1}\mbox{-}\bar{\bm{3}}]+\bar{b}\bar{b})$ & $\sigma(g\gamma\to \Xi_{bb}[bb,{^1S_0}\mbox{-}\bm{6}]+\bar{b}\bar{b})$ & Total \\
				\hline
				Pb-Pb      & 5.52  & 60.0 nb   & 4.81 nb  & 129.62 nb  \\
				\hline
				Xe-Xe      & 5.86  & 19.2 nb   & 1.56 nb  & 41.52 nb  \\
				\hline
				Kr-Kr      & 6.46  & 6.96 nb   & 0.57 nb  & 15.06 nb  \\
				\hline
				Ar-Ar      & 6.3   & 920 pb    & 75.6 pb  & 1991 pb  \\
				\hline
				Ca-Ca      & 7.0   & 1299 pb   & 107 pb   & 2812 pb  \\
				\hline
				O-O        & 7.0   & 99.5 pb   & 8.27 pb  & 215.5 pb  \\
				\hline
				p-Pb       & 8.8   & 523 (58.8) pb   & 42.9 (5.11) pb   & 629.8 pb  \\
				\hline
				p-p        & 14    & 454 fb    & 39.8 fb  & 987.6 fb   \\
				\hline
				\hline
				Pb-Pb      & 39.4  & 546 nb    & 48.1 nb  & 1188 nb  \\
				\hline
				p-Pb       & 62.8  & 4.17 (0.30) nb  & 0.37 (0.027) nb  & 4.867 nb    \\
				\hline
				p-p        & 100   & 2245 fb   & 206 fb   & 4902 fb  \\
				\hline		
			\end{tabular}
		}
	\end{center}
	\label{Tabgr2Xibb}
\end{table} 

We also study the production cross sections for $\Xi_{bc/bb}$ via various elastic photoproduction at HL-LHC and FCC, the numerical results for each spin-color states are listed in TABLE \ref{Tabgr2Xibc} and \ref{Tabgr2Xibb}. As the heavy constituent for $bc$-quark are different, the exchange asymmetry for identical particle is not hold, hence all diquark structures for spin-color states in $bc[{^3S_1}\mbox{-}\bar{\bm{3}}],\ bc[{^3S_1}\mbox{-}\bm{6}],\ bc[^1S_0\mbox{-}\bar{\bm{3}}]$, and $bc[^1S_0\mbox{-}\bm{6}]$ will contribute in a comparable level. At the HL-LHC, $10^{4}\sim 10^{6}$ $\Xi_{bc}$ could be produced, and increase to $2\times10^{6}\sim 7\times10^{7}$ for FCC, led to open possibility for phenomenological researches. For the $\Xi_{bb}$, the production cross sections are further suppressed, only 648 $\Xi_{bb}$ could be produced via Pb-Pb collision. Hence the phenomenological investigation for $\Xi_{bb}$ through ion-ion elastic photoproduction may be not feasible. 

\section{SUMMARY AND CONCLUSIONS}
In this work, we investigate the doubly heavy baryon and tetraquark production via photon-photon and photon-gluon fusion with various ultraperipheral ion-ion collisions in the framework of NRQCD factorization formalism. Two cc(bb)-diquark configurations, $[cc(bb),{^3S_1}\mbox{-}\bar{\bm{3}}]$ and $[cc(bb),{^1S_0}\mbox{-}\bm{6}]$, are considered for $\Xi_{cc(bb)}$ ($T_{cc(bb)}$) production; four bc-diquark configurations, $[bc,{^3S_1}\mbox{-}\bar{\bm{3}}]$, $[bc,{^3S_1}\mbox{-}\bm{6}]$, $[bc,{^1S_0}\mbox{-}\bar{\bm{3}}]$ and $[bc,{^1S_0}\mbox{-}\bm{6}]$, are taken into account for $\Xi_{bc}$ production. The cross sections for each diquark configurations and the total cross sections versus transverse momentum and rapidity at the HL-LHC and FCC are given. The cross section for doubly charmed tetraquark $T_{cc}$ is also estimated.

Numerical results show that the cross section for $[cc,{^3S_1}\mbox{-}\bar{\bm{3}}]$ diquark is around $20\sim30$ times of $[cc,{^1S_0}\mbox{-}\bm{6}]$ diquark for photon-photon fusion, the ratio decreases to $11\sim12$ for photon-gluon fusion. Based on designed luminosities of each ion-ion collision at the HL-LHC and FCC, a considerable number of doubly charmed baryon and tetraquark can be expected. Due to the event topologies for ultraperipheral collision are vary clear, the background from various QCD interactions can be suppressed, hence the detailed experimental investigation for $\Xi_{cc}$ and $T_{cc}$ are feasible. The transverse momentum and rapidity distributions are given. The productions for $\Xi_{bc/bb}$ are also discussed, leaving only slightly possibility for observing $\Xi_{bc}$ through photon-gluon fusion at the FCC. 

\vspace{1.4cm} {\bf Acknowledgments}
This work is supported by the National Natural Science Foundation of China (NSFC) under the Grants
Nos. 12275185 and 12335002.

\end{document}